\documentclass[letterpaper]{article} 
\usepackage{arxiv}

\usepackage{times}  
\usepackage{helvet}  
\usepackage{courier}  
\usepackage[hyphens]{url}  
\usepackage{graphicx} 
\usepackage{natbib}  
\usepackage{caption} 
\usepackage{algorithm}
\usepackage{algorithmic}

\usepackage{newfloat}
\usepackage{listings}

\usepackage{amsmath} 
\usepackage{array}
\usepackage{multirow}
\usepackage{booktabs} 
\usepackage{makecell}
\usepackage{xcolor}
\usepackage{float}

\DeclareCaptionStyle{ruled}{labelfont=normalfont,labelsep=colon,strut=off} 
\floatstyle{ruled}
\newfloat{listing}{tb}{lst}{}
\floatname{listing}{Listing}
\title{Bayesian Sensitivity Analyses for Policy Evaluation with Difference-in-Differences under Violations of Parallel Trends}

\author{
Seong Woo Han\textsuperscript{\rm 1},
Nandita Mitra\textsuperscript{\rm 2},
Gary Hettinger\textsuperscript{\rm 2,3}\textsuperscript{\dag},
Arman Oganisian\textsuperscript{\rm 4}\textsuperscript{\dag}
}

\affiliations{
\textsuperscript{\rm 1}Department of Computer and Information Science, University of Pennsylvania\\
\textsuperscript{\rm 2}Division of Biostatistics, University of Pennsylvania\\
\textsuperscript{\rm 3}Department of Population Health, New York University\\
\textsuperscript{\rm 4}Department of Biostatistics, Brown University\\
seonghan@seas.upenn.edu, nanditam@pennmedicine.upenn.edu, gary.hettinger@nyulangone.org, arman\_oganisian@brown.edu

}

\usepackage{bibentry}

\begin{document}

\maketitle
\begingroup
\renewcommand\thefootnote{\dag}
\footnotetext{Co-senior authors}
\endgroup

\begin{abstract}
Violations of the parallel trends assumption pose significant challenges for causal inference in difference-in-differences (DiD) studies, especially in policy evaluations where pre-treatment dynamics and external shocks may bias estimates. In this work, we propose a Bayesian DiD framework to allow us to estimate the effect of policies when parallel trends is violated. To address potential deviations from the parallel trends assumption, we introduce a formal sensitivity parameter representing the extent of the violation, specify an autoregressive AR(1) prior on this term to robustly model temporal correlation, and explore a range of prior specifications -- including fixed, fully Bayesian, and empirical Bayes (EB) approaches calibrated from pre-treatment data. By systematically comparing posterior treatment effect estimates across prior configurations when evaluating Philadelphia’s sweetened beverage tax using Baltimore as a control, we show how Bayesian sensitivity analyses support robust and interpretable policy conclusions under violations of parallel trends.

\end{abstract}

\section{Introduction}

To evaluate the causal impact of public policy, practitioners commonly use DiD designs, which compare outcome trends between treated and control groups while adjusting for time-invariant differences \cite{angrist2009mostly}. A key identifying assumption in DiD is that, in the absence of treatment, the two groups would have followed parallel trends. This assumption is often difficult to justify in real-world policy evaluations, where complex pre-treatment dynamics, time-varying confounding, or unobserved effect modifiers may violate the parallel trends condition \cite{zeldow2021confounding}. For instance, a DiD analysis of Philadelphia’s sweetened beverage tax using Baltimore as a control found a 51\% decline in city-level beverage sales following the tax \cite{roberto2019association}, but such estimates may be sensitive to deviations from the assumed counterfactual trend.

A common approach for evaluating the plausibility of parallel trends is pre-trends testing, which checks whether treated and control groups followed similar outcome trajectories before treatment. While widely used, such tests are underpowered in short panels or with noisy outcomes and do not guarantee the trends would have remained parallel post-treatment. Further, as noted by \cite{roth2022pretest}, pre-tests can also introduce selective inference when only favorable results are reported. Moreover, failing to detect a pre-treatment difference does not confirm the assumption holds after treatment. To address potential deviations more directly, researchers have proposed adjusting for covariates that may drive differential trends. Methods such as semi-parametric propensity score-based or double robust DiD estimators \cite{abadie2005semiparametric, sant2020doubly} and Bayesian regression models \cite{normington2019bayesian} aim to reduce bias by modeling outcomes and weighting on observed confounders. However, their validity depends on correctly specified covariates, and unmeasured confounding in observational settings can still produce uncorrected violations. Given these limitations, sensitivity analyses have become essential tools for assessing the robustness of DiD estimates to violations of the parallel trends assumption. Rather than treating the assumption as binary, these methods formally relax it and examine how inferences shift under bounded or parameterized deviations from the ideal.

Recent advances \cite{rambachan2023more, manski2018right, keele2019patterns} have introduced partial identification frameworks that yield credible bounds on treatment effects when parallel trends may not hold. These frameworks forgo point identification and instead define sets of plausible treatment effects by bounding deviations from parallel trends. For example, \cite{rambachan2023more} allow for post-treatment violations constrained by smoothness (e.g., second-difference restrictions) or magnitude (e.g., bounding violations by a fixed multiple of pre-treatment differences). These non-parametric assumptions are agnostic to the violation’s functional form and do not rely on stochastic modeling. Rather than estimate a violation process directly, they specify a feasible deviation set based on observed pre-trends, leading to partially identified treatment effects and valid confidence intervals.

While these methods yield robust inference, they often rely on closed-form constraints—such as bounding post-treatment deviations by pre-treatment magnitudes or enforcing smoothness—that can be overly conservative and limit their ability to capture data-driven violations. Alternatively, Bayesian approaches to sensitivity analysis may offer greater flexibility by incorporating uncertainty and prior knowledge directly into the model. For example,\cite{kwon2024empirical} propose an approach to relax the parallel trends assumption by modeling deviations as an AR(1) prior in which each period's deviation depends on the previous one and random noise. However, their model lacks a mean-shift parameter, constraining the long-run mean of deviations to zero and thus excluding systematic upward or downward trends. While \citet{kwon2024empirical} adopt an EB framework, their implementations differ across applications. Their application to \citet{benzarti2019really} calibrates the AR(1) parameters externally using estimates from \citet{mcgahan1999persistence}, rather than estimating them directly from their own pre-treatment data. For the \citet{lovenheim2019long} analysis, they adopt a simplified random walk with drift, modeling period-to-period changes as normally distributed around a constant mean, with parameters estimated from pre-treatment data via maximum likelihood. This formulation lacks mean reversion and imposes a weaker temporal structure than AR(1), meaning parallel trends deviations can accumulate unchecked over time, thereby reducing the model’s ability to distinguish temporary fluctuations from sustained shifts. While conceptually Bayesian, these models do not conduct formal sensitivity analyses—they do not systematically vary parameters or assess how treatment effect estimates change under alternative specifications.

In this study, we improve upon existing approaches by assessing treatment effects under both the parallel trends assumption and structured deviations from it, using a Bayesian framework that models and quantifies potential violations through sensitivity parameters. Focusing on the Philadelphia sweetened beverage tax policy and using Baltimore as a control \cite{roberto2019association}, we analyze beverage sales data from January 2016 to December 2017 to account for potential deviations in pre-treatment trends due to market differences and external factors. To address this, we implement Bayesian sensitivity analysis with three prior structures, including an AR(1) model for temporal deviations and an EB approach calibrated from the observed pre-treatment data. This enables a data-driven evaluation of how prior assumptions influence treatment effect estimates. Our results show the flexibility of Bayesian methods in handling violations of parallel trends and highlight the importance of sensitivity analysis for robust causal inference in policy evaluation.


\section*{DiD identification under parallel trends violations}

To formalize the identification strategy and how it changes under violations of parallel trends, we begin by outlining the potential outcomes framework for DiD.

\paragraph{Setting} Suppose we observe a sequence of periods \( t \in \{0, 1, \dots, T\} \), where periods \( t < g \) represent pre-policy periods, and \( t \geq g \) represent post-policy periods \cite{callaway2021difference}. A unit is either exposed to the policy from time \( g \) onwards (\( A = 1 \)) or remains unexposed (\( A = 0 \)). Define the potential outcome at time \( t \), had a unit been exposed to policy status \( a \), as \( Y^a(t) \). We are interested in estimating the average treatment effect on the treated (ATT), which is the average difference in post-policy outcomes between exposed and unexposed units, among those exposed. We express this as:

\begin{equation}
\Psi_t = E[Y^1(t) - Y^0(t) \mid A = 1]
\end{equation}

Using the linearity of expectation, this becomes \( \Psi_t = E[Y^1(t) \mid A = 1] - E[Y^0(t) \mid A = 1] \). The challenge arises because we cannot directly observe \( E[Y^0(t) \mid A = 1] \), as it represents the counterfactual outcome of exposed units had they not been exposed to the policy. 

\paragraph{Parallel trends assumption} To address this issue, the parallel trends assumption is invoked, particularly in multi-period settings. This assumption states that, in the absence of treatment, the outcome trend among treated units would have followed the same trend as that among untreated units:

\begin{equation}
\begin{aligned}
E[Y^0(t) - Y^0(t-1) \mid A = 1] = \\
 E[Y^0(t) - Y^0(t-1) \mid A = 0] 
& \quad \text{for } t \geq g
\end{aligned}
\end{equation}

This implies that we can express the counterfactual mean outcome of treated units in period \( t \) in terms of pre-policy outcomes and trends observed among untreated units:

\begin{equation}
\begin{aligned}
E[Y^0(t) \mid A = 1] = E[Y^0(t-1) \mid A = 1] &+ \\
\left( E[Y^0(t) - Y^0(t-1) \mid A = 0] \right)
\end{aligned}
\end{equation}

Under the consistency assumption, which states that observed outcomes equal potential outcomes under the received treatment (i.e., \( Y(t) = Y^a(t) \) if \( A = a \)), we can substitute into the ATT definition for \( \Psi_t \):

\begin{equation}
\begin{aligned}
\Psi_t = \left( E[Y(t) \mid A = 1] - E[Y(g-1) \mid A = 1] \right) - \\
\left( E[Y(t) \mid A = 0] - E[Y(g-1) \mid A = 0] \right)
\end{aligned}
\end{equation}

This equation identifies the ATT under the parallel trends assumption using observed outcome trajectories across groups. While identification only requires pre-treatment data from the immediate time before treatment ($t=g-1$), earlier pre-treatment data can still benefit estimation by improving statistical efficiency and enabling the assessment of pre-treatment trend alignment.






\paragraph{Violations of parallel trends assumption} The parallel trends assumption posits that, in the absence of treatment, the outcome trends for treated and untreated groups would evolve in the same way. However, when this assumption is violated, the expected trend in the treated group may deviate from that of the untreated group. Formally, this deviation is captured by the equation:

\begin{equation}
\begin{aligned}
E[Y^0(t) - Y^0(t - 1) \mid A = 1] = \\
E[Y^0(t) - Y^0(t - 1) \mid A = 0] + \xi_t
\end{aligned}
\end{equation}

where $\xi_t$ quantifies the deviation from parallel trends between time $t-1$ and $t$, for each $t \geq g$. If $\xi_t = 0$ for all $t$, then parallel trends hold, meaning that the untreated and treated groups follow the same expected trajectory. Any nonzero $\xi_t$ introduces a deviation from parallel trends, where the sign of $\xi_t$ determines the direction of this deviation. Given this modified parallel trends assumption, we can express the counterfactual mean outcome of the treated group at any period $t \geq g$ as:

\begin{equation}
\begin{aligned}
E[Y^0(t) \mid A = 1] = E[Y^0(t - 1) \mid A = 1] &+ \\
E[Y^0(t) - Y^0(t - 1) \mid A = 0] + \xi_t
\end{aligned}
\end{equation}

This recursive equation describes how the expected counterfactual outcome at each period $t$ is derived from its previous value, incorporating the expected change in the untreated group along with the deviation term $\xi_t$. To generalize this to an accumulated form, we apply this recursion iteratively from the treatment onset period $g$ to $t$, summing over all incremental changes. Expanding iteratively, we obtain the telescoping sum:

\begin{equation}
\begin{aligned}
E[Y^0(t) \mid A = 1] = E[Y^0(g-1) \mid A = 1] + \\
\sum_{s=g}^{t} (E[Y^0(s) \mid A = 1] - E[Y^0(s-1) \mid A = 1])
\end{aligned}
\end{equation}

Substituting the modified parallel trends assumption into the summation, we replace each term with its equivalent expression:

\begin{equation}
\begin{aligned}
E[Y^0(s) \mid A = 1] - E[Y^0(s-1) \mid A = 1] &= \\
E[Y^0(s) - Y^0(s-1) \mid A = 0] + \xi_s
\end{aligned}
\end{equation}

This leads to:

\begin{equation}
\begin{aligned}
E[Y^0(t) \mid A = 1] = E[Y^0(g-1) \mid A = 1] + \\ \sum_{s=g}^{t} \Big( E[Y^0(s) - Y^0(s-1) \mid A = 0] + \xi_s \Big)
\end{aligned}
\end{equation}

Applying the linearity of expectation, we simplify the summation over the untreated group’s expected changes:

\begin{equation}
\begin{aligned}
E[Y^0(t) \mid A = 1] = E[Y^0(g-1) \mid A = 1] + \\ E[Y^0(t) \mid A = 0] - E[Y^0(g-1) \mid A = 0] + \sum_{s=g}^{t} \xi_s
\end{aligned}
\end{equation}

Finally, the consistency assumption allows us to replace remaining potential outcomes with observed outcomes. We equate the treated and control group's potential outcomes at time \( t \) with their observed outcomes, replacing \( E[Y^1(t) \mid A = 1] \), \( E[Y^0(t) \mid A = 0] \), and \( E[Y^0(g{-}1) \mid A = 0] \) with their observed values. Under the arrow of time or no anticipation assumption, future intervention does not affect past outcomes, allowing us to equate \( E[Y^0(g{-}1) \mid A = 1] \) with \( E[Y(g{-}1) \mid A = 1] \), yielding:

\begin{equation}
\begin{aligned}
E[Y(t) \mid A = 1] = E[Y(g-1) \mid A = 1] + \\
E[Y(t) \mid A = 0] - E[Y(g-1) \mid A = 0] + \sum_{s=g}^{t} \xi_s
\end{aligned}
\end{equation}
This final equation captures the cumulative deviation of the treated group's counterfactual outcomes from the untreated group's expected trend, accounting for both structural trend differences and accumulated deviations $\xi_s$. The summation term reflects how deviations at each time $s$ accumulate, leading to a total shift in expectations by time $t$.


\section*{Sensitivity models for parallel trends violation}

We now introduce modeling strategies for estimating treatment effects under both parallel trends and its violations. 

\paragraph{Two-way fixed effects (TWFE) estimation} Each expectation above can be estimated using a multi-period TWFE model, which is specified as follows:

\begin{equation}
E[Y_i(t) \mid A = a] = \theta_t + \gamma + \beta \mathbf{1}_{\{t \geq g, A = 1\}}
\end{equation}

In this model, \( Y_i(t) \) represents the outcome variable for unit \( i \) at time \( t \). The term \( \theta_t \) captures time-specific fixed effects, which account for any time-varying factors common across all units, effectively controlling for temporal trends unrelated to the policy. The term \( \gamma \) represents group-specific fixed effects, which account for unobserved characteristics specific to treated (\( A = 1 \)) or untreated (\( A = 0 \)) groups that remain constant over time. The parameter \( \beta \) is associated with the time-dependent treatment status indicator \( \mathbf{1}_{\{t \geq g, A = 1\}} \), and it represents the treatment effect for exposed units in periods \( t \geq g \), which corresponds to the ATT in this multi-period setting. The TWFE approach is advantageous because it isolates the policy’s effect by controlling for both time-specific and group-specific confounding factors. This helps ensure that any detected effect is due to the policy intervention rather than other time-related or group-specific influences. Posterior inference for \( \Psi_t \) can then be achieved by estimating the posterior distribution of \( \beta \) using a Bayesian linear regression model.

\paragraph{AR(1) prior on deviations from parallel trends}

To allow for deviations from the parallel trends assumption, we introduce an AR(1) prior on the sequence of post-treatment violation terms \(\{\xi_g, \xi_{g+1}, \dots, \xi_t\}\). The AR(1) prior provides a structured way to model how deviations evolve over time, ensuring that each period's violation builds incrementally upon the prior period's deviation. Specifically, for each period \( s = g, g+1, \dots, t \), we define:

\begin{equation}
\xi_s = \eta(1 - \rho) + \rho\,\xi_{s-1} + \sigma\,\varepsilon_s 
\label{eq:ar1}
\end{equation}

Here, \(\varepsilon_s \sim \text{iid } N(0, 1)\), and \(\eta\), \(\rho\), and \(\sigma\) are the parameters of the process. This structure is denoted as \(\{\xi_s\}_{s=g}^t \sim gAR1(\eta, \rho, \sigma)\). The mean of this AR(1) is \(E[\xi_s] = \eta\). The variance is given by \(V[\xi_s] = \frac{\sigma^2}{1 - \rho^2}\). The parameter \(-1 < \rho < 1\) is the autoregressive parameter that governs the dependence on the state, and \(\sigma\) is the standard deviation of the noise term. Under this specification, each deviation $\xi_s$ is incrementally dependent on the deviation of the preceding period $\xi_{s-1}$. Consequently, the cumulative deviation term for periods $g$ to $t$ given by $\sum_{s=g}^{t}\xi_s$ captures structured incremental departures at each step, as each deviation systematically builds on the previous one. This AR(1) prior offers several advantages for modeling longitudinal deviations in DiD settings. By explicitly accounting for autocorrelation, it ensures that deviations are not treated as independent across time but evolve gradually, reflecting realistic temporal dependencies common in panel or repeated-measures data. This not only enables more robust inference in the presence of serially correlated violations but also regularizes the trajectory of \(\xi_s\), preventing overfitting to local noise. Specifically, our approach differs from \citet{kwon2024empirical}, who model the deviation process as an autoregressive structure with no mean shift. In their formulation, the deviation term (analogous to our $\xi_s$) follows $\xi_s = \rho \xi_{s-1} + \sigma \varepsilon_s$, thereby assuming the deviations are centered around zero over time. This corresponds to an AR(1) process with zero long-run mean. In contrast, we explicitly incorporate a mean-shift parameter $\eta$ in the AR(1) formulation, $\xi_s = \eta(1 - \rho) + \rho \xi_{s-1} + \sigma \varepsilon_s$, allowing for persistent directional shifts in the counterfactual trend. Moreover, rather than calibrating parameters externally as done in their application, we estimate $(\eta, \rho, \sigma)$ empirically from pre-treatment data, yielding a fully data-adaptive prior.

\section*{Sensitivity analysis and parameter variants on Philadelphia beverage tax (PBT) data}

We now describe the dataset and implementation details used in our Bayesian analysis of the PBT. We then outline the prior configurations used to model deviations from parallel trends.

\paragraph{Application to PBT dataset} 

Beverage price and sales data were sourced from the market research firm Information Resources Inc (IRI), which collects data from major US retailers \cite{muth2016understanding}. The PBT, implemented on January 1, 2017, imposed a 1.5 cent per ounce tax on the distribution of both sugar-sweetened and artificially sweetened beverages. This policy aimed to reduce consumption of sugary drinks and raise revenue for public programs. For this analysis, we used beverage sales data from January 1, 2016, to December 31, 2017, collected from supermarkets and pharmacies in Philadelphia and Baltimore. Observations were recorded at 26 time points, indexed as \( t = 1, \dots, 26 \), where each period represents a four-week aggregation of sales. The first 13 periods (\( t = 1, \dots, 13 \)) correspond to the pre-tax periods in 2016, and the remaining 13 periods (\( t = 14, \dots, 26 \)) correspond to the post-tax periods in 2017. This structure enables consistent temporal comparisons across years. Store and beverage categorization, as well as price and sales aggregations, were conducted as described in \cite{roberto2019association}.

For this study, we classified stores into two groups: supermarkets, which include grocery stores and mass merchandisers, and pharmacies, recognizing that consumer purchasing behavior may differ between these store types. In addition, a treated indicator was added to identify observations from treated locations, specifically Philadelphia during the post-tax period, defined by the indicator variable \( \mathbf{1}(t > 13) \). This categorization facilitates comparative analysis of the tax's effect on beverage sales between Philadelphia and Baltimore, providing a structured approach to assess the impact across time points.

\begin{table}[t]
\centering
\scriptsize
\begin{tabular}{c ccc ccc} 
\toprule
\multirow{2.5}{*}{\shortstack{AR(1) \\ models}} & \multicolumn{3}{c}{Supermarket} & \multicolumn{3}{c}{Pharmacy} \\
\cmidrule(lr){2-4} \cmidrule(lr){5-7}
 & $\sigma$ & $\rho$ & $\eta$ & $\sigma$ & $\rho$ & $\eta$ \\
\midrule
Fixed-1 & 0.001 & 0.95 & U(0.1, 0.9) & 0.001 & 0.95 & U(0.1, 0.9) \\
Fixed-2 & 1 & 0.95 & U(0.1, 0.9) & 1 & 0.95 & U(0.1, 0.9) \\
Fixed-3 & 5 & 0.95 & U(0.1, 0.9) & 5 & 0.95 & U(0.1, 0.9) \\
\midrule 
Fully-1 & HN(1) & B(2,2) & U(0.1, 0.9) & HN(1) & B(2,2) & U(0.1, 0.9) \\
Fully-2 & HN(2) & B(2,2) & U(0.1, 0.9) & HN(2) & B(2,2) & U(0.1, 0.9) \\
Fully-3 & HN(5) & B(2,2) & U(0.1, 0.9) & HN(5) & B(2,2) & U(0.1, 0.9) \\
\midrule
EB-1 & 0.166 & 0.371 & 1.60 & 0.340 & 0.785 & 1.64 \\
EB-2 & 0.166 & 0.742 & 1.60 & 0.340 & 1.57 & 1.64 \\
EB-3 & 0.166 & 1.113 & 1.60 & 0.340 & 2.36 & 1.64 \\
\bottomrule
\end{tabular}
\caption{Configurations of AR(1) models with parameter configurations across supermarket and pharmacy data. U = Uniform, HN(k) = HalfNormal(k), B = Beta.}
\label{table:ar_models}
\end{table}

\begin{figure*}[t]
    \centering
    \includegraphics[width=1.0\textwidth]{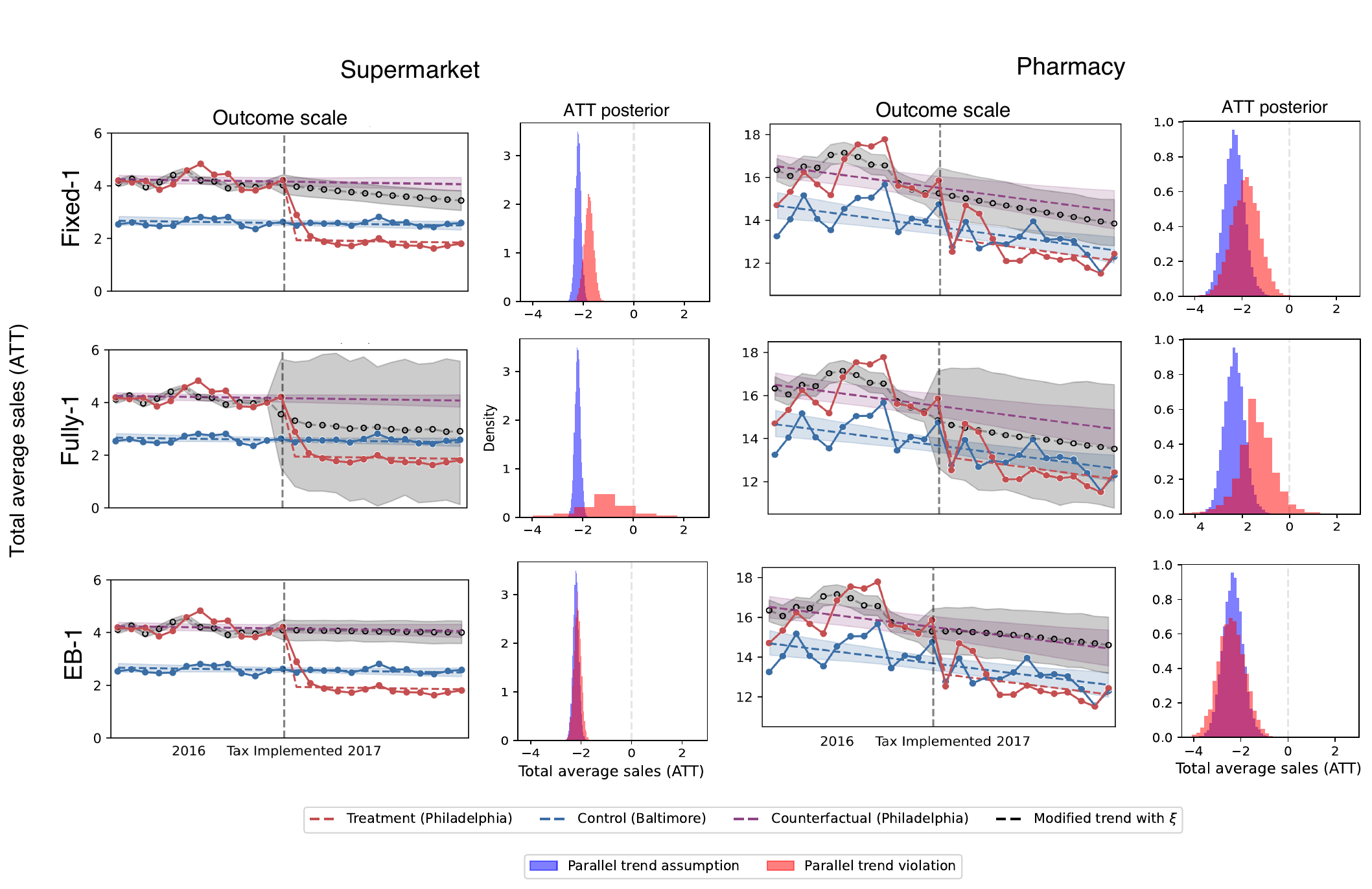} 
    \caption{Outcome scale and ATT posterior for Supermarket and Pharmacy datasets. The outcome plots (left panels) show the observed treatment trend (Philadelphia), observed control trend (Baltimore), counterfactual trend assuming no treatment, and modified trend incorporating an AR(1) prior with sensitivity parameter $\xi$. The ATT posterior plots (right panels) compare the difference in outcomes between the treatment and counterfactual trends under the parallel trends assumption, and the difference between the treatment and modified trends with $\xi$ under parallel trend violation. The histograms reflect the posterior distribution of the total ATT accumulated over the post-treatment period, up to the final time point. Each row corresponds to Fixed-1, Fully-1, and EB-1 models from top to bottom respectively, as defined in Table~\ref{table:ar_models}. The analysis highlights the impact of increasing sensitivity parameters on outcome trends and ATT posteriors.}
    \label{fig:figure1} 
\end{figure*}

\paragraph{Hyperparameter specification strategies}

To assess the sensitivity of our results to assumptions about deviations from parallel trends, we specify three distinct strategies for selecting the hyperparameters of the AR(1) prior—specifically, the standard deviation $\sigma$, autoregressive coefficient $\rho$, and long-run mean, $\eta$. These strategies correspond to three groups of models, fixed (Fixed-1, Fixed-2, Fixed-3), fully Bayesian (Fully-1, Fully-2, Fully-3), and empirical Bayes (EB-1, EB-2, EB-3), each applied separately to the supermarket and pharmacy data. The complete parameter configurations are listed in Table~\ref{table:ar_models}.

In the fixed models, we fix the autoregressive parameter at \( \rho = 0.95 \), a value close to 1 that preserves strong temporal dependence while still allowing the AR(1) to include a non-zero long-run mean \( \eta \). We draw \( \eta \sim \text{Uniform}(0.1, 0.9) \) and vary the standard deviation \( \sigma \) across 0.001 (Fixed-1), 1 (Fixed-2), and 5 (Fixed-3). A very small \( \sigma \) tightly constrains deviations toward the mean, approximating the parallel trends assumption, while larger values permit increasingly flexible departures. This fixed setup simplifies the sensitivity analysis by isolating the role of deviation scale, but it omits posterior uncertainty in the autoregressive structure and may overstate confidence  as it does not incorporate uncertainty in \( \rho \) and \( \sigma \).

The fully Bayesian models incorporate fully Bayesian priors to account for uncertainty in the autoregressive structure. We place a \( \text{Beta}(2,2) \) prior on \( \rho \), reflecting moderate belief in positive autocorrelation while avoiding extremes near 0 or 1. For the noise scale, we assign \( \sigma \sim \text{HalfNormal}(k) \) with \( k \in \{1, 2, 5\} \) across the models Fully-1 to Fully-3, respectively, allowing increasing flexibility in deviation magnitude. The long-run mean \( \eta \) remains drawn from \( \text{Uniform}(0.1, 0.9) \). This approach enables the model to learn both the smoothness and volatility of deviations from the data, producing more realistic posterior uncertainty. However, its effectiveness may still depend on the informativeness of the priors, especially when the number of post-treatment periods is short.

The EB models estimate \( \rho \), \( \sigma \), and \( \eta \) from pre-treatment trends using an EB procedure (see next subsection). EB-1 uses the raw EB estimates, while EB-2 and EB-3 apply multiplicative scaling to \( \sigma \) by factors of 2 and 5, respectively, to evaluate robustness to prior regularization strength. This data-driven approach allows the prior to adapt to observed trends, but may be sensitive to noise or bias in the pre-treatment period.

\begin{figure*}[t]
  \centering
  \includegraphics[width=1.0\textwidth]{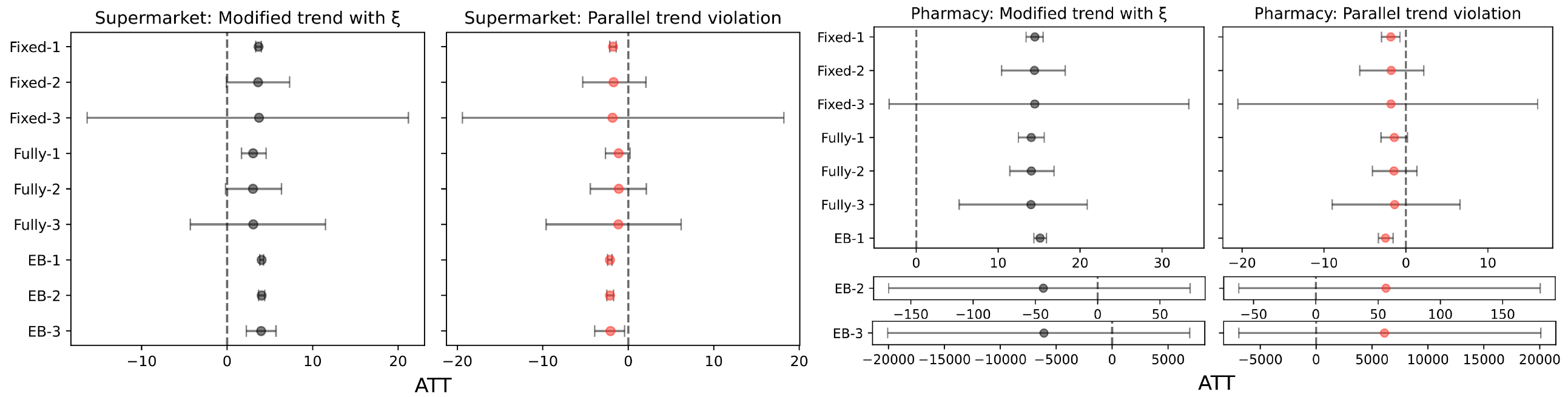}
  \caption{
  Posterior $\mu$ and 95\% CI for ATT estimates across all AR(1) configurations from Table~\ref{table:ar_models}, with left panels summarizing the modified trend with parallel trend violation $\xi$ from outcome scale plots and right panels summarizing the parallel trend violation condition from ATT posterior histograms, both shown in Figure~\ref{fig:figure1}. For the pharmacy dataset, EB-2 and EB-3 are plotted on separate horizontal scales due to extreme estimates driven by nonstationary $\rho$ values ($|\rho| \geq 1$), which amplify deviations and inflate CI.
  }
  \label{fig:att_summary}
\end{figure*}

\paragraph{EB estimation of AR(1) parameters}

To enable data-adaptive regularization, we empirically estimate parameters for the AR(1) model \eqref{eq:ar1} directly from observed pre-treatment trends. Let \( X_t \) denote the sequence of estimated violation terms \( \xi_t \) over the pre-treatment periods. We rewrite the model as $X_t = c + \rho X_{t-1} + \sigma \epsilon_t$, where \( \epsilon_t \sim N(0, 1) \), and \( c = \eta (1 - \rho) \). This representation makes the model linear in \( c \) and \( \rho \). Using ordinary least squares, we estimate

\begin{equation}
\begin{bmatrix} c \\ \rho \end{bmatrix} = (\mathbf{Z}^\top \mathbf{Z})^{-1} \mathbf{Z}^\top \mathbf{X}
\end{equation}

where \(\mathbf{Z}\) includes a column of ones and lagged values \( X_{t-1} \). This procedure minimizes the residual sum of squares, yielding an estimate of \( \rho \) as the influence of \( X_{t-1} \) on \( X_t \). With estimates for \( c \) and \( \rho \), we derive \( \eta \) as \( \eta = \frac{c}{1 - \rho} \), providing an estimate of the long-term mean based on the short-term dynamics captured by the model. To estimate \( \sigma \), which represents the standard deviation of the noise term, we use the residuals from the model, calculated as \( \hat{\epsilon}_t = X_t - (c + \rho X_{t-1}) \). These residuals approximate the noise component \( \sigma \epsilon_t \) in \eqref{eq:ar1}. The empirical variance of these residuals provides an estimate for \( \sigma^2 \):

\begin{equation}
\hat{\sigma}^2 = \frac{1}{n-2} \sum_{t=2}^n \left(X_t - (c + \rho X_{t-1})\right)^2
\end{equation}

Taking the square root of \( \hat{\sigma}^2 \) gives \( \hat{\sigma} \), which serves as an empirical measure of the variability captured by \( \sigma \) in \eqref{eq:ar1}. 

\paragraph{Implementation setup} The models were implemented in PyMC (version 4.0) \cite{salvatier2016probabilistic}. The default sampler is the no-U-turn sampler, an adaptive form of Hamiltonian Monte Carlo \cite{betancourt2017, hoffman2014}.  The default number of chains is four, and the default runs 1000 warmup iterations (for burn-in and adaptation) and 1000 sampling iterations.  All sampling runs ended with split-$\widehat{R}$ values less than 1.01 for all parameters, indicating consistency with convergence to approximate stationarity \cite{gelman2013bayesian}.

\section*{Results of sensitivity analysis}
To evaluate how assumptions about violations of the parallel trends assumption influence treatment effect estimates, we perform a series of Bayesian sensitivity analyses using the AR(1) prior configurations described in Table~\ref{table:ar_models}. While nine total model configurations were used across supermarket and pharmacy datasets, we focus on Fixed-1, Fully-1, and EB-1 models as representative trends from each hyperparameter strategy in Table~\ref{table:ar_models}. The full results for all models are provided in Figure~\ref{fig:att_summary}.

\paragraph{Outcome scale and ATT posterior}

Each panel in Figure~\ref{fig:figure1} consists of two components, the outcome scale (left) and the ATT posterior (right). The outcome scale plot shows observed sales for treatment and control groups, alongside two modeled counterfactuals: one based on extrapolating pre-treatment trends under the parallel trends assumption, and another adjusted by an autoregressive deviation process governed by a sensitivity parameter $\xi$ which allows for flexible violations. The ATT posterior histogram summarizes the estimated treatment effects under each assumption, calculated as the difference between the observed treatment outcome and its respective counterfactual.

In the supermarket panels of Figure~\ref{fig:figure1}, Fixed-1, Fully-1, and EB-1 models highlight distinct behaviors driven by their respective deviation dynamics. Fixed-1 exhibits a gradual downward shift in its counterfactual, reflecting the accumulation of modest but persistent deviations over time. Fully-1 shows even greater divergence and wider uncertainty because its priors allow the counterfactual trend to drift more freely. Unlike Fixed-1 and EB-1, which are tightly anchored to pre-treatment trends, Fully-1 places non-informative priors, $\rho \sim \text{Beta}(2,2)$ and $\sigma \sim \text{HN}(1)$, which do not constrain the model to closely follow the pre-tax trajectory, allowing a more flexible fit that permits substantial downward trends after treatment. In contrast, EB-1 remains closely aligned with the extrapolated trend, as its empirically estimated parameters yield near-stationarity and minimal cumulative deviation. These differences in counterfactual behavior directly influence the ATT posteriors. Fully-1 yields the most attenuated effects under violations, while EB-1 demonstrates the greatest robustness. Similar patterns are observed in the pharmacy dataset.

\paragraph*{Summary of posterior estimates}
Figure~\ref{fig:att_summary} summarizes the posterior $\mu$ and 95\% CI for the ATT estimates from Figure~\ref{fig:figure1}, with the left panels corresponding to the modified trend with $\xi$ from the outcome scale plot, and the right panels correspodning to the parellal trend violoation condition from the ATT posterior histogram, across all nine configurations defined in Table~\ref{table:ar_models}. As expected in sensitivity analysis, increasing model flexibility—through higher $\sigma$ (Fixed-1 to Fixed-3), wider priors (Fully-1 to Fully-3), or EB scaling (EB-1 to EB-3)—leads to greater posterior uncertainty and shifts in the mean estimates. The fixed models show a clear gradient: as $\sigma$ increases, so does the variance in both modified trends and ATT under violation. The fully Bayesian models mirror this, with HalfNormal priors capturing uncertainty more realistically but still exhibiting sensitivity to prior scale. In the EB models, posterior sensitivity is generally low, especially in EB-1, where all AR(1) parameters are directly calibrated from pre-treatment data. However, EB-2 and especially EB-3 yield unstable results in the Pharmacy dataset due to excessively high estimated values of $\rho$ (1.57 and 2.36, respectively). Since AR(1) processes with $|\rho| \geq 1$ become nonstationary, these cases amplify small fluctuations, leading to extreme values and inflated credible intervals. In practice, posterior estimates with $|\rho| \geq 1$ should be treated as a diagnostic signal that the model may be overfitting short-term fluctuations. To prevent nonstationary drift, we recommend using priors such as Beta distributions over $\rho \in (0,1)$ that enforce or softly encourage stationarity, especially when pre-treatment data are limited.

\begin{figure}[ht]
  \centering
  \includegraphics[width=0.33\textwidth]{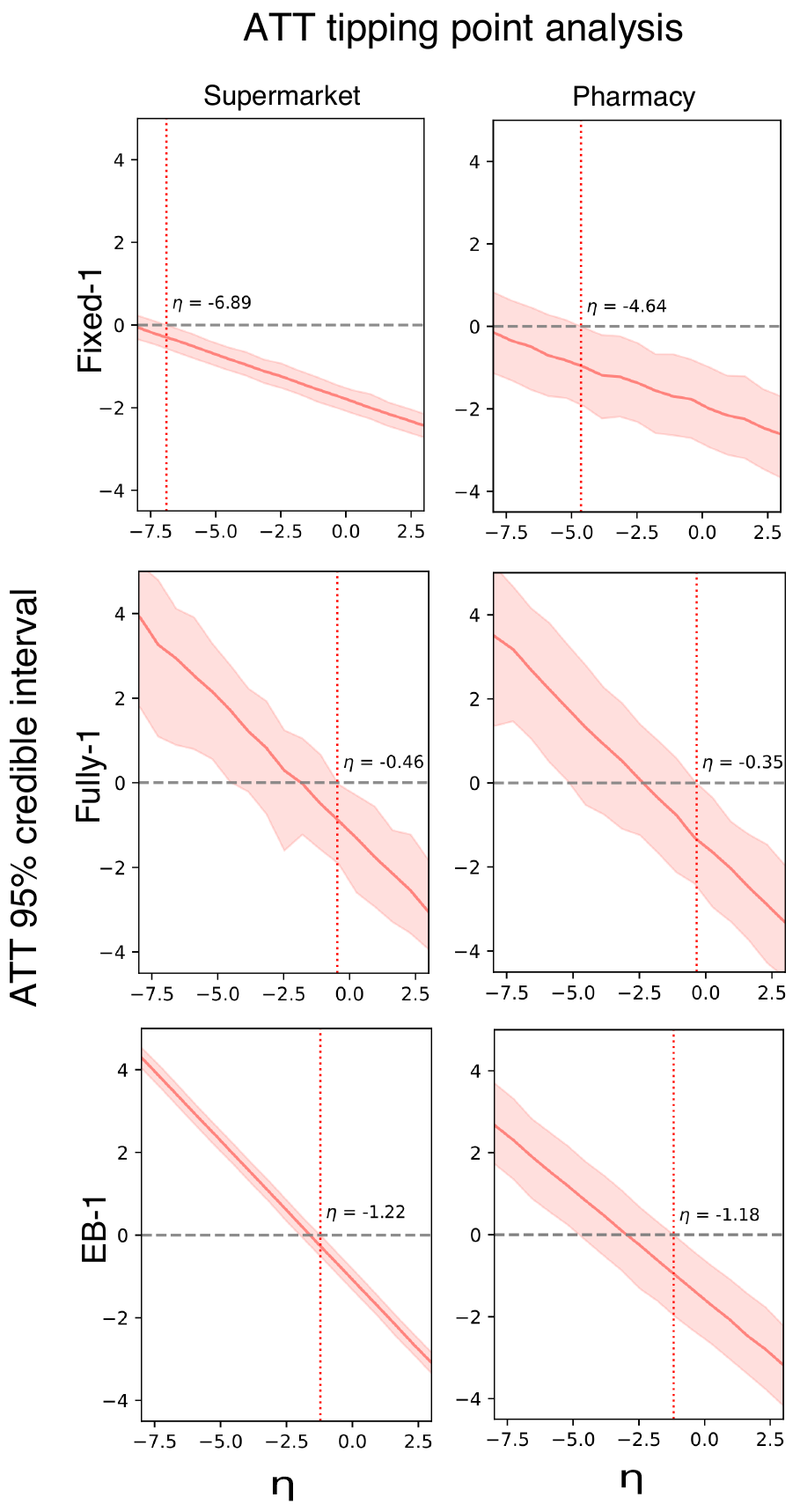}
  \caption{
    Tipping point analysis assessing the robustness of ATT estimates to violations of parallel trends. Each panel shows the posterior $\mu$ and 95\% CI of the ATT across values of the sensitivity parameter $\eta$, which controls the degree of deviation via an AR(1) structure. The dotted red line marks the $\eta$ value where the upper bound of the 95\% CI crosses zero—indicating the tipping point beyond which the ATT is no longer significantly negative. Results are shown for three representative models (Fixed-1, Fully-2, and EB-1) across supermarket (left) and pharmacy (right) datasets.}
  \label{fig:att_tipping}
\end{figure}

\paragraph*{Tipping point analysis for sensitivity to trend violations}
To assess the robustness of our treatment effect estimates to violations of the parallel trends assumption, we examine how large the sensitivity parameter $\eta$ must be before the upper bound of the ATT’s 95\% CI includes zero. Since $\eta = 0$ corresponds to strict parallel trends, we vary $\eta$ in both directions—emphasizing negative values that shift the ATT toward the null—to test how easily the effect could be attenuated. Figure~\ref{fig:att_tipping} shows posterior $\mu$ and 95\% CIs for the ATT across $\eta$ values under Fixed-1, Fully-1, and EB-1 models. In our AR(1) framework, as $\eta$ introduces additive shifts in sales growth rates, exponentiating these deviations yields multiplicative changes in sales volume, enabling fold-change interpretation relative to the parallel trend.

In our default specifications, the supermarket and pharmacy models assume $\eta = 1.6$ and $\eta = 1.64$, respectively (Table~\ref{table:ar_models}). Based on the PBT dataset \cite{roberto2019association}, average beverage sales were 4.85 million ounces per 4-week period for supermarkets and 0.16 million ounces for pharmacies. For supermarkets under Fully-1 model, the tipping point occurs at $\eta = -0.46$, which implies that untreated sales would have been approximately 58\% higher than the counterfactual under parallel trends. Applying this increase to the original average of 4.85 million ounces yields an additional 2.81 million ounces per store. Since one can of soda is 12 ounces, this corresponds to roughly 234,000 extra cans per 4-week period. For pharmacies under Fully-1 model, the tipping point at $\eta = -0.35$ implies a 42\% increase in untreated sales. When applied to the baseline volume of 0.16 million ounces, this equates to approximately 67,000 more ounces, or about 5,600 additional cans. EB-1 model reaches its tipping point at $\eta = -1.2$, requiring a 3.32-fold increase in sales. This equates to 11.25 million extra ounces (938,000 cans) for supermarkets and 371,000 ounces (30,900 cans) for pharmacies. Fixed-1 model reaches their tipping points at $\eta = -6.89$ and $\eta = -4.64$, requiring 980-fold and 104-fold increases—over 4.7 billion additional ounces (395 million cans) for supermarkets and 16.4 million ounces (1.37 million cans) for pharmacies. These results show that Fixed-1 model demands large counterfactual increases to overturn the effect, EB-1 model requires substantial but unlikely deviations, and Fully-1 model is the most sensitive to modest departures from parallel trends.

\section*{Discussion}

Our Bayesian sensitivity analyses, enabled by an AR(1) prior on the violation process, underscore the importance of modeling deviations from the parallel trends assumption in DiD studies. However, this analysis focuses on a two-region comparison between Philadelphia and Baltimore, which may miss spillover effects or cross-border shopping that influence nearby beverage sales. Expanding the analysis to include indirectly affected regions and covariates for regional heterogeneity could offer a more complete view of the policy’s broader impact, though this adds complexity when using a TWFE model \cite{sant2020doubly, hettinger2025doubly}. Additionally, the EB method’s reliance on pre-treatment trends may reduce its reliability when such data are limited or unrepresentative. Overall, our findings suggest that credible inferences about policy effects remain possible even when parallel trends are violated, with sensitivity analyses offering a flexible framework that allows practitioners to gauge how conclusions vary with different modeling choices and pre-treatment information.

\nobibliography*
\bibliography{aaai2026}

\end{document}